\documentclass[12pt]{article}

\usepackage{siunitx}
\usepackage{times}
\usepackage[version=4]{mhchem}
\usepackage{graphicx}

\usepackage{soul,color}

\title{Ultra-thin van der Waals crystals as semiconductor quantum wells} 

\author
{Johanna Zultak,$^{1,2}$ Samuel Magorrian,$^{1,2}$ Maciej Koperski,$^{1,2}$\\ 
Alistair Garner,$^3$ Matthew Hamer,$^{1,2}$ Endre T\'ov\'ari,$^{1,2}$\\ 
Kostya S. Novoselov,$^{1,2}$ Alexander Zhukov,$^{1,2}$ Yichao Zou,$^{2,3}$\\ Neil R. Wilson,$^4$ Sarah J. Haigh,$^{2,3}$ Andrey Kretinin,$^{2,3}$\\ Vladimir I. Fal'ko,$^{1,2,5*}$ 
Roman Gorbachev$^{1,2,5*}$\\
\\
\normalsize{$^{1}$School of Physics and Astronomy, University of Manchester, Oxford Road,}\\
\normalsize{ Manchester, M13 9PL, UK}\\
\normalsize{$^{2}$National Graphene Institute, University of Manchester, Oxford Road,}\\
\normalsize{ Manchester, M13 9PL, UK}\\
\normalsize{$^{3}$School of Materials, University of Manchester, Oxford Road,}\\
\normalsize{ Manchester, M13 9PL, UK}\\
\normalsize{$^{4}$Department of Physics, University of Warwick, Coventry, CV4 7AL, UK}\\
\normalsize{$^{5}$Henry Royce Institute for Advanced Materials, University of Manchester,}\\
\normalsize{Oxford Road, Manchester, M13 9PL, UK}\\
\\
\normalsize{ E-mail: vladimir.falko@manchester.ac.uk, roman@manchester.ac.uk.}
}

\date{}

\begin{document} 
\maketitle 

\section*{Abstract}
Control over the electronic spectrum at low energy is at the heart of the functioning of modern advanced electronics: high electron mobility transistors, semiconductor and Capasso terahertz lasers, and many others. Most of those devices rely on the meticulous engineering of the size quantization of electrons in quantum wells. This avenue, however, hasn’t been explored in the case of 2D materials. Here we transfer this concept onto the van der Waals heterostructures which utilize few-layers films of InSe as quantum wells. The precise control over the energy of the subbands and their uniformity guarantees extremely high quality of the electronic transport in such systems. Using novel tunnelling and light emitting devices, for the first time we reveal the full subbands structure by studying resonance features in the tunnelling current, photoabsorption and light emission. In the future, these systems will allow development of elementary blocks for atomically thin infrared and THz light sources based on intersubband optical transitions in few-layer films of van der Waals materials.  

\section*{Introduction}
Van der Waals crystals provide a platform for layer-by-layer material design which has expanded into a large multidisciplinary field during the last decade \cite{Geim2013,Novoselov2016,Britnell2013,Li2016,Akinwande2017,Bandurin2017,Raja2017,Zhang2018}. A broad variety of electronic and optoelectronic applications have been described and proof-of-concept devices have been reported \cite{Novoselov2016,Akinwande2017}. Along with high quantum yield \cite{Britnell2013}, compact design, and mechanical flexibility \cite{Akinwande2017}, one of the attractive features of two dimensional semiconductors (2DS) is their band-gap tunability. Due to the change in quantum confinement with the thickness, the primary optical transition can be tuned over a broad range (\SI{> 1}{\eV} \cite{Li2016,Bandurin2017,Zhang2018} ), and further fine adjustment is available through changes in dielectric environment \cite{Raja2017} and transverse electric fields \cite{Haastrup2016}. 

While the majority of 2DS demonstrate primary optical transitions in visible and near-infrared (IR) ranges, their intersubband transitions hold a largely unexplored potential to expand their optical activity further into IR and THz. Depending upon the choice of material, number of layers and doping (n- or p-type), these transitions densely populate a region from \SI{\sim 0.8}{eV} for bilayers to \SI{0.05}{eV} for films $\sim$10 layers and can be utilized for optoelectronic devices \cite{Magorrian2018,Ruiz-Tijerina2018}. The experimental studies of higher energy subbands remained elusive until recently, when the first experimental observation of the intersubband transitions in few-layer \ce{WSe2} has been demonstrated \cite{Schmidt2018} using near-field THz absorption spectroscopy. While THz spectroscopy gives access to direct measurements of some intersubband transitions, it only works in a small energy window and requires a specific doping of the materials studied. At the same time, angle-resolved photoemission spectroscopy (ARPES) only shows filled (valence) subbands and has diminished energy resolution when applied to micrometre-size crystals \cite{Hamer2019}. Here, we present a promising experimental approach that enables a comprehensive experimental study of the full subband structure of atomically thin 2DS, namely resonant tunneling spectroscopy combined with photoluminescence excitation (PLE) measurements. When applied to InSe films, this method enables us to map the subbands on both conduction and valence band sides of the spectrum in agreement with theoretical band structure modeling and available ARPES data, as well as trace the analogy between atomically thin films and quantum wells in conventional semiconductors \cite{Klitzing1987}.

\section*{Results and discussion}
\begin{figure}
 \centering
 \includegraphics[width=8.8cm]{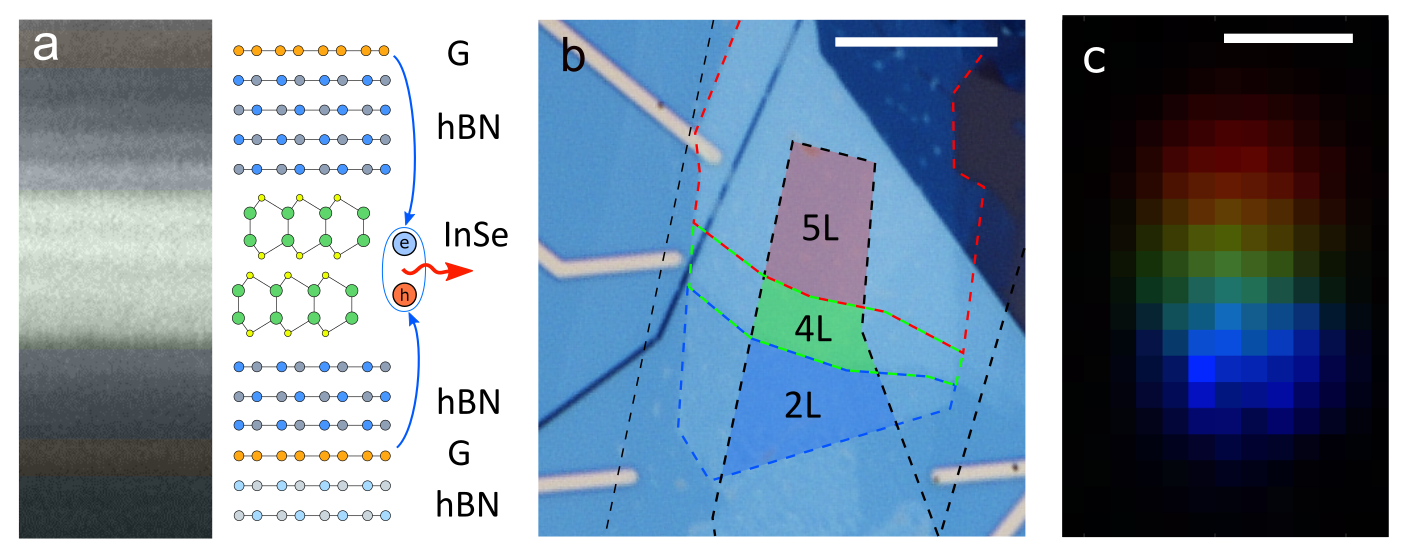}
 \caption{\label{fig:I} \textbf{InSe tunneling devices.} (\textbf{a}) Cross-sectional high-angle annular dark-field scanning transmission electron image of a typical LED device with matching layer schematic on the right. (\textbf{b}) Optical micrograph of a device used to study resonant tunneling through InSe subbands. Graphene is outlined in black, 2L InSe in 
blue, 4L in green and 5L in red. (\textbf{c}) Artificially colored EL map of the device shown in (b): blue corresponding to EL around \SI{1.98}{\eV}, green \SI{1.50}{\eV} and red \SI{1.35}{\eV} with sampling bandwidth of \SI{1}{\meV} (detailed EL spectra can be found in SI). The scale bars are \SI{10}{\um}.}
\end{figure}

Indium selenide is a new material in the 2DS family. It has a layer-dependent bandgap spanning from \SI{1.2}{eV} for bulk to almost \SI{3.0}{eV} for a monolayer \cite{Mudd2013,Lei2014,Bandurin2017} and features a crossover \cite{Zolyomi2014,Magorrian2016,Hamer2019} from weakly indirect (in mono-, bi- and tri-layer) to direct character (in crystals thicker than 4 layers). Its high crystal quality has recently been demonstrated, resulting in exceptionally high electron mobility \cite{Bandurin2017} making it a promising avenue for developing atomically thin nanoelectronics \cite{Hamer2018}. For resonant tunneling measurements we build van der Waals heterostructures, where exfoliated few-layer $\gamma$-InSe is sandwiched between hexagonal boron nitride (hBN) crystals to provide tunneling barriers, and we employ graphene layers as source and drain contacts on the top and the bottom of the device illustrated in Fig.\ref{fig:I}a. The thickness of hBN was selected to be 4-3 layers to allow for large bias voltages without sample overheating. These structures have been produced using the dry-stacking method in an argon environment \cite{Frisenda2018} to provide atomically clean and sharp interfaces, and to avoid InSe degradation \cite{Cao2015,Rooney2017}. The assembled stacks were deposited on an oxidized silicon wafer and contacts were defined using electron beam lithography (see supplementary information). 

\begin{figure}
 \centering
 \includegraphics[width=\textwidth]{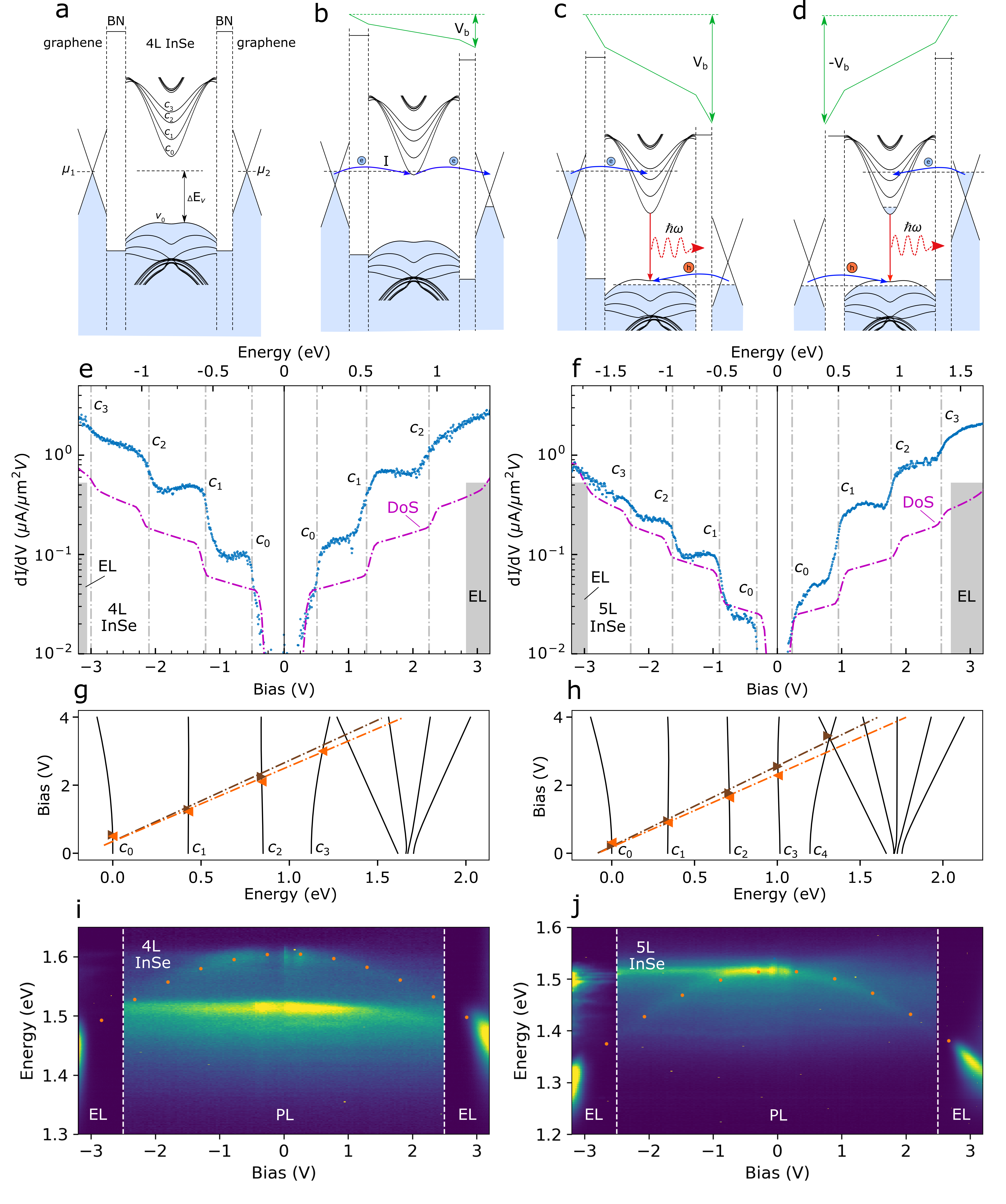}
 \caption{\label{fig:II} \textbf{Resonant tunneling spectroscopy of InSe films at T=4.2K.} Band alignment in 4L InSe device, for unbiased \textbf{(a)}, weakly biased \textbf{(b)} and strongly biased in forward \textbf{(c)} and reversed \textbf{(d)} configurations. $dI/dV$ (blue) and density of states {of the conduction bands }from tight binding model (purple, arb. u.) for 4L \textbf{(e)} and 5L \textbf{(f)} InSe film, with the number of layers established by AFM topography. Energy scale along the top axis was found using the EL onset values. Evolution of the conduction subbands with bias for 4L (\textbf{g}) and 5L (\textbf{h}) InSe  (higher energy bands are also shown). Symbols indicate the bias at which the steps were measured in (e) and (f) for forward (brown) and reverse (orange) directions. Evolution of PL and EL with bias for 4L \textbf{(i)} and 5L \textbf{(j)} InSe. The orange dotted line indicates electrostatic calculations of the band gap reduction with electric field, non-dispersive line near \SI{1.5}{eV} is due to hydrocarbon contamination located outside the sample.}
\end{figure}

An optical micrograph of a typical device consisting of three InSe regions $N=$ 2, 4 and 5 layers thick is shown in Fig.\ref{fig:I}b. Each of these terraces has separate graphene contacts allowing them to have bias applied individually and control the position of chemical potentials $\mu_{1}$ and $\mu_{2}$ in source and drain graphene layers, respectively. To understand vertical charge transport in this system, we have reconstructed its band diagram using a tight-binding-calculated InSe band-structure \cite{Magorrian2016,Bandurin2017} and the relative band alignment measured in our recent ARPES studies \cite{Hamer2019} using InSe exfoliated from the same bulk ingot. Here, the graphene Dirac point was near the NL-InSe {(N layers-InSe)} valence band edge. The resulting band diagram is shown in Fig.\ref{fig:II} for unbiased (a), weakly biased (b), strongly biased in forward (c) and reverse configuration (d) 4L-InSe LED device. 

All these anticipated vertical transport regimes can be recognized in the  experimentally measured {tunneling current derivative} curves shown in Fig.\ref{fig:II}e and f. As soon as a small bias \SI{\sim 0.3}{V} is applied, {$dI/dV_b$} shows a sharp increase (for 4L-InSe (e) and 5L-InSe (f)), indicating that $\mu_1$ crosses the edge of the first InSe conduction subband, in agreement with the band diagram in Fig.\ref{fig:II}b. {Upon further increase in $V_{b}$, $dI/dV_b$ displays a series of steps.} We attribute this behavior to the step-like increase in the density of states of InSe when $\mu_{1}$ crosses the conduction subbands $c_\mathrm{i}$. Eventually, $\mu_{2}$ of the drain graphene reaches the top subband $v_0$ of the valence band, electrons and holes injected into the InSe from opposite sides recombine, producing a bright electroluminescence (EL) signal, shown in the intensity map Fig.\ref{fig:I}c \cite{Withers2015}. At this point the tunneling current reaches \SI{\sim 1}{\micro A/\micro m^2}, and becomes unstable due to the contribution from recombination process.

Using bias values at the onset of EL $V_{b}^{th}$ (indicated as grey shaded areas in Fig.\ref{fig:II}e,f) we can determine the rate of change of InSe potential with bias, as $\alpha={\Delta E_{v}}/{V_{b}^{th}}$, where $\Delta E_v$ is offset between Dirac point in graphene and InSe valence band edge at zero bias \cite{Hamer2019}. The EL signal is observed to have slight asymmetry for forward and reverse bias voltages: e.g. the thresholds of EL for 4L InSe LED are $V_{b}=$\SI{2.83}{V} and \SI{-3.02}{V}, due to the different hBN barrier thicknesses (4L and 3L) on either side. Therefore, the relation between $V_b$ and energy in the band structure of the crystal is slightly different for forward and reverse bias (top axis in Fig.\ref{fig:II}e,f). Using the experimentally determined values of parameter $\alpha$ for both devices, we are able to compare the experimentally measured $dI/dV_b$ with the theoretically calculated density of electron states (dashed purple line). We note that the linear density of states (DoS) in graphene makes the assignment of observed steps unambiguous, as its only feature, the Dirac point, is located near zero bias where tunneling current is not measurable. This enables us to determine the energies of subband edges experimentally.  

For completeness, in the theoretical analysis we took into account the shifts of both conduction and valence subbands caused by the out-of-plane electric field due to the applied bias as, 
\begin{equation}
	E = \frac{V_b}{d_{\mathrm{InSe}}+\frac{\varepsilon_{\mathrm{InSe}}}{\varepsilon_{\mathrm{BN}}}d_{\mathrm{BN}}},
\end{equation}
where $d_{\mathrm{InSe}}$ and $d_{\mathrm{BN}}$ are the respective total thicknesses of the InSe and hBN in the stack, and $\varepsilon_{\mathrm{InSe}}/{\varepsilon_{\mathrm{BN}}}\approx 4.9$ is the ratio between the out-of-plane dielectric constants of InSe and hBN. This leads to a clearly pronounced quadratic red shift of the lowest subband$^\prime$s energy seen in Fig.\ref{fig:II}g,h, similar to previously the observed quantum confined Stark effect in \ce{MoS2} monolayers \cite{Roch2017a}. At the same time, subbands in the middle of the spectrum ($c_1, c_2$ for 4L and $c_1, c_2, c_3$ for 5L) remain almost constant, whereas the highest subband is blue shifted. The DoS calculations shown in Fig. 2e,f take into account the bias dependence of the band structure and the resulting subband onsets match closely (within \SI{40}{\meV}) with the step-like $dI/dV_b$ features observed in our experiment (showed as grey dashed lines at the maximum derivative).

The calculated subband shifts for both conduction and valence bands agree well with the shift in the photoluminescence (PL) energy shown in Fig.\ref{fig:II}i,j. The observed evolution of EL with bias also accurately follows the trend predicted for PL and at $V_b\approx$ \SI{3}{V} is shifted by almost \SI{200}{meV} since the $c_{0}$ and $v_{0}$ bands move in opposite directions. It is also clearly non-symmetric, with an almost double EL linewidth for the negative biases, which can be explained with partial filling of the $c_0$ subband when the hBN barrier of the drain is thicker than that of the source. This also agrees with a much faster increase in EL intensity for negative $V_b$, indicating that the recombination rate is primarily determined by a steeper rise in the DoS of the flatter valence band. 

\begin{figure}
 \centering
 \includegraphics[width=\textwidth]{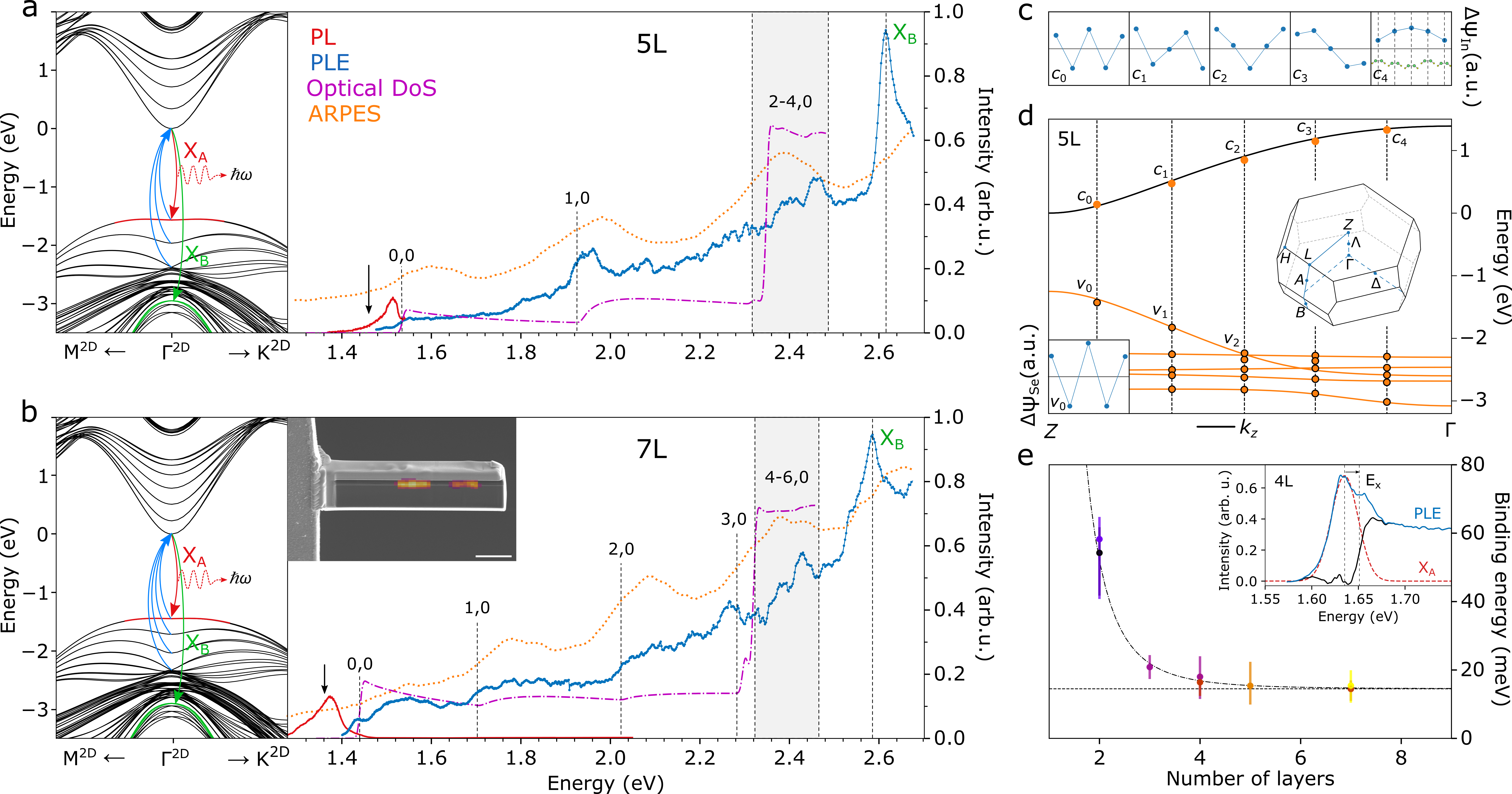}
 \caption{\label{fig:PLE} \textbf{PLE of 5L and 7L InSe at T=4.2K.} \textbf{(a,b)} Left - 2D band structure of 5 and 7L InSe around $\Gamma$ computed using DFT-parametrized tight-binding dispersion with spin-orbit interaction taken into account. Right - PL (red), PLE measured at the black arrow (blue) and projected optical DoS (purple) of lamellae for 5 and 7L InSe. Relevant optical transitions from $v_i$ to $c_j$ are labeled as \textit{i,j}. Orange dotted line shows ARPES intensity around $\Gamma$-point \cite{Hamer2019}, plotted to match $v_0$-$c_0$ optical transition in PLE with the first valence subband in ARPES. Inset shows scanning electron micrograph of the lamella overlaid with its PL map sampled at \SI{1.3\pm0.02}{\eV}, scale bar is \SI{5}{\micro m}. \textbf{(c)} Amplitude $\Delta \psi_{\mathrm{In}} = \psi(\mathrm{In_1})-\psi(\mathrm{In_2})$, of In $S$-orbital  contribution to the subbands $c_n$  across the layers for 5L-InSe from tight-binding model. \textbf{(d)} Solid lines - out-of-plane dispersion of bulk InSe at in-plane $\Gamma$ point. Orange dots - $\Gamma$-point subbands for 5L InSe at $k_z$ given by quantization of momentum in quantum well picture (dashed lines). Insets: center – 3D Brillouin zone of InSe, bottom left - amplitude $\Delta \psi_{\mathrm{Se}} = \psi(\mathrm{Se_1})-\psi(\mathrm{Se_2})$, of Se $P_z$-orbital contribution to top valence subband. \textbf{(e)} Exciton binding energy ($E_X$) extracted from PLE (see inset) as a function of number of layers in the film. Colors indicate different samples. The dashed line is displayed as a guide to the eye.} 
\end{figure}

To illustrate the behavior of the subbands in the context of few-layer InSe as a quantum well, we show in Fig.\ref{fig:PLE}d the $k_z$ dispersion of the bulk crystal together with the subband energies for 5-layer InSe at the in-plane $\Gamma$ point. The 5-layer subband energies can be qualitatively matched with the bulk bands, quantized at 5 discrete $k_z$ momenta as is familiar in the analysis of subband spectra in semiconductor quantum wells \cite{Lage1991}. We further illustrate the quantum well behavior of the subbands in Fig.\ref{fig:PLE}c by plotting conduction band wavefunction coefficients for $S$ orbitals on the (antisymmetric) indium pair in each layer, showing how the subband with the lowest effective $k_z$ has no nodes, with the number of nodes increasing with $k_z$.

Finally, we probed the subband states of InSe films using PLE, with the typical quasi-absorption spectra shown in Fig.\ref{fig:PLE}a,b for $N=5$ and $7$ layers. Here, InSe films were encapsulated in thicker (\SI{>20}{\um}) hBN without graphene contacts. The optical excitation from the lowest conduction band to the highest valence subband is mostly coupled to light polarized perpendicular to the plane of 2D crystal due to the intralayer electric dipole moment \cite{Magorrian2016}. To provide both in- and out-of-plane polarization of photons, we have extracted a cross sectional block, with an hBN/InSe/hBN area of \SI{2}{\micro m} x \SI{20}{\micro m} and a depth of \SI{4}{\micro m} using a focused ion beam instrument (see supplementary information) and rotated it \ang{90} such that the viewing direction is along the basal plane of the InSe film. To interpret the PLE data we note that the material studied was lightly p-doped, so that the lowest-energy emission{, $X_A$,} used for detection involved equilibrium (rather than photoexcited) holes in the subband $v_0$ (dominated by the P$_z$ orbitals of Se \cite{Magorrian2016}). Taking into account that selection rules favour radiative recombination to the subband $v_0$ from the conduction band subband $c_0$, which have the same distribution of amplitudes across the layers (Fig.\ref{fig:PLE}c,d), we assume that PLE detects excitation primarily into $c_0$, so that we compare the observed PLE spectra with the optical density of states projected onto the lowest subband on the conduction band side in the final states. This optical density of states takes into account transitions from all subbands {$v_{0,...,N-1}$} on the valence band side  as well as highly optically active transitions \cite{Magorrian2016} from the subbands generated by the deeper band based on P$_{x,y}$ orbitals of Se atoms ({$X_B$}) that are mixed with the top valence band by spin-orbit coupling.\cite{Bandurin2017} 

The theoretically calculated projected optical DoS is plotted in Fig.\ref{fig:PLE}a,b (purple dashed lines). It accounts for absorption of both out- and in-plane polarised photons: the latter is caused by the spin-flip interband transitions generated by spin-orbit mixing of  P$_z$ orbitals of Se with the P$_{x,y}$ orbital (marked in {green} in Fig.\ref{fig:PLE}a,inset)\cite{Magorrian2016}. While such mixing is weak for the states in the highest subband $v_0$, it is enhanced for the deeper subbands, which are closer to the P$_{x,y}$ band on the energy scale, and this is reflected by the overall increase of optical DoS towards higher photon energies. The photon energy thresholds for each transition $v_n \rightarrow c_0$, determined from the spectra of standing waves subbands in the film, Fig.\ref{fig:PLE}c, are indicated (as ``n,0") in Fig.\ref{fig:PLE}a,b, and coincides with the features in the observed PLE spectra. For comparison, on the same plot we show micro-ARPES spectra measured on 5L and 7L films of InSe, which reflect the subband structure of the valence band in these films with noticeably broader features.

Also, on all samples studied, detailed inspection of the low energy onset of the PLE spectra have shown a noticeable excitonic peak followed by a plateau due to the free particle absorption. Measuring the difference between the peak energy (red line in the inset of Fig.\ref{fig:PLE}e) and the onset of continuum absorption (black line) \cite{kittel2004}, we estimate the layer-dependent exciton binding energies as shown in Fig.\ref{fig:PLE}e. The sharp increase of the binding energy for thin samples agrees with those typically seen in other 2DS \cite{Wang2017}, while for thick flakes ($N > 5$) it closely matches (within \SI{1}{meV}) that of bulk InSe \cite{Camassel1978}.


By this we achieve comprehensive understanding of the entire subband spectrum of few-layer atomically thin films of InSe, including its evolution due to the quantum confined Stark effect, enabled by the combined resonant tunneling and photoluminescence excitation spectroscopy approach. Having been exemplified here for InSe, this approach can be extended to a variety of 2D systems, including hetero- and homo-bilayers of transition metal dichalcogenides with a view to detect features of moiré
 superlattice minibands \cite{Alexeev2019}. For optoelectronic applications, electrical injection of carriers in vertical devices is a possible route towards creating infrared and terahertz light sources that employ intersubband transitions. The above results demonstrate that, given appropriate doping, one can control carrier injection into any designated subband of an atomically thin crystal to enable such emitters. 

\section*{Methods}
\paragraph*{Device fabrication}
For all devices Bridgman-grown bulk rhombohedral $\gamma$-InSe \cite{Hamer2019} crystals were mechanically exfoliated onto PPC-coated (polypropylene carbonate) silicon wafers and crystals displaying large uniform terraces were identified using optical microscopy. These flakes were picked up with graphene/hBN stack carried on a PMMA membrane \cite{Frisenda2018} followed by a second thin hBN crystal and a graphene monolayer. Finally, the assembled stack was released onto a thick ($\sim$ \SI{50}{nm}) hBN exfoliated on an oxidized silicon wafer. The exfoliation and transfer of thin InSe took place in an argon filled glovebox to protect it from degradation \cite{Hamer2018}. Both graphene crystals were then contacted using electron beam lithography followed by CHF${_3}$ etching and Cr/Au (\SI{1.5}{\nm}/\SI{50}{\nm}) deposition. Using an additional e-beam lithography step, the top graphene layer was divided into separate electrodes following outlines of the InSe terraces using Ar/O$_2$ plasma etching. 

For the measurements of PL and PLE, devices without graphene have been made to prevent the emission quenching \cite{Gaudreau2013,Raja2016}. As before, InSe was encapsulated between hBN crystals and then covered with an additional thick hBN layer (\SI{>200}{\nm}) followed by Cr/Au (\SI{1}{\nm}/\SI{80}{\nm}) to prevent Ga$^{+}$ ion damage of the InSe during the milling process. Using FEI Helios dual-beam focused ion beam (FIB) / scanning electron microscope (SEM), an additional layer of platinum (\SI{\sim1}{\um}) was deposited over the selected region to provide further protection. Ga+ FIB milling was used to define the lamella at \SI{30}{kV}, with current \SI{7}{\nA} to remove bulk of the material and current \SI{1}{nA} to trim the lamella to its correct size. An OmniProbe micromanipulator was used to extract the lamella, rotate it by \ang{90}, and position it on an OMICRON transmission electron microscopy (TEM) grid. After milling, the damaged edges of the 2D stack were removed by FIB polishing using decreasing acceleration voltages ( \SI{5}{kV}, \SI{47}{pA} and \SI{2}{kV}, \SI{24}{pA}). The final thickness of the specimen was \SI{\sim2}{\um}.

\paragraph*{Optical measurements}  
All optical measurements have been performed at \SI{4}{K} using an AttoDry100 cryostat and a AttoCFM I inset from Attocube. The illumination power on the sample was kept around \SI{0.4}{\mW} and the spot size was \SI{\sim 2}{\um}.  For the excitation, a supercontinuum white light laser (Fianium WhiteLase WL-SC-400-15-PP) combined with a  Contrast filter (LLTF SR-VIS-HP8) have been used resulting in accurate wavelength control between \SI{400}{\nm} and \SI{1000}{\nm} with a linewidth of \SI{\sim 2.5}{\nm}. For PLE measurements, the laser power was stabilized before the sample using an RD40-UV laser power controller from Brockton Electro-Optics Corp. The PL/PLE spectra were recorder using spectrometer diffraction  grating (Princeton Instruments Acton Spectrapro  SP-2500i with \SI{300}{g/\mm}) and nitrogen-cooled CCD camera (Princeton Instruments Pylon PIX-100BRX). 

\paragraph*{Tunneling Spectroscopy}
The tunneling measurements were carried out at \SI{4.2}{\K} by applying a DC bias and measuring the current between the source and drain graphene electrodes using a Keithley 2614B source-meter.

\paragraph*{Tight-binding model}
Details concerning the spin-orbit coupling, the calculation of the density of states and the absorption spectra can be found in the Supplementary Materials.

\paragraph*{Cross-sectional STEM}
For high resolution scanning transmission electron microscope (STEM) imaging the FIB cross section received further thinning with a Helios Dual beam FIB-SEM, using sequentially lower milling currents of \SI{30}{kV}, \SI{16}{kV}, \SI{5}{kV} and \SI{2}{kV} for ion beam milling and polishing. A probe side aberration corrected FEI Titan G2 80-\SI{200}{kV} was used with a probe convergence angle of \SI{21}{mrad}, a HAADF inner angle of \SI{48}{mrad} and a probe current of \SI{\approx 80}{pA}. To ensure the electron probe was parallel to the basal planes, the cross-sectional FIB sample was aligned to the relevant Kikuchi bands of the Si substrate and the 2D crystals. 

\paragraph*{Data availability}
Additional data related to this paper may be requested from the authors.

\section*{Acknowledgments}
We acknowledge support from EPSRC grants EP/N509565/1, EP/P01139X/1, EP/N010345/1, EP/P01139X/1 and EP/L01548X/1 along with the CDT Graphene-NOWNANO, and the EPSRC Doctoral Prize Fellowship. In addition, we acknowledge support from the  European Graphene Flagship Project (785219), European Quantum Technology Flagship (820378) 2D-SIPC, Marie Skłodowska Curie Scheme (751883), ERC Synergy Grant Hetero2D, ERC Starter grant EvoluTEM, the ARCHER National UK Supercomputer RAP Project e547, Royal Society, US Army Research Office (W911NF-16-1-0279) and Lloyd Register Foundation Nanotechnology grant. We acknowledge support from Dr Alexei Barinov at Spectromicroscopy beamline, ELETTRA synchrotron, with the ARPES data.

\section*{Author contributions}
V.I.F. and R.G. conceived the study; J.Z. fabricated devices with help from M.H.; J.Z. and M.K. performed measurements with help of E.T. and A.K.; J.Z. analyzed experimental data; S.J.H. and Y.Z. performed electron imaging; S.M. and V.I.F. provided theory used for the interpretation of the experiments; A.G. and A.Z. performed lamella preparation; N.W. provided ARPES data for comparison; V.I.F., R.G. and J.Z. wrote the manuscript with the help of S.M., A.K. and K.S.N; 




\end{document}